\documentclass{article} 
\usepackage{iclr2025_delta,times}

\usepackage{hyperref}
\usepackage{url}
\usepackage{times}
\usepackage{soul}
\usepackage{url}
\usepackage[utf8]{inputenc}
\usepackage[small]{caption}
\usepackage{graphicx}
\usepackage{amsmath}
\usepackage{amsthm}
\usepackage{booktabs}
\usepackage[switch]{lineno}

\newcommand{\lightgray}[1]{{\color{lightgray} {#1}}}

\usepackage{graphicx} 
\usepackage{verbatim}
\usepackage{amsmath,amssymb,amsfonts,bm}
\usepackage{bbm}
\usepackage{enumitem}
\usepackage{amsthm}
\usepackage{algpseudocode}
\usepackage{textcomp}
\usepackage{float}
\usepackage[font=small]{caption}
\usepackage{xcolor}
\usepackage{wrapfig}
\usepackage{hyperref}
\usepackage[font=footnotesize,labelformat=simple]{subcaption}
\usepackage[protrusion=true,expansion=true]{microtype}
\pdfoutput=1

\usepackage{caption}
\usepackage{multirow}
\usepackage{graphicx}
\usepackage{slashbox}
\usepackage{amsthm}

\usepackage[font=footnotesize,labelformat=simple]{subcaption}
\everymath{\displaystyle}

\title{Physics-Informed Generative Approaches for Wireless Channel Modeling}

\author{Satyavrat Wagle\\
Elmore School of Electrical and Computer Engineering, Purdue University\\
\texttt{wagles@purdue.edu}
\AND
Akshay Malhotra, Shahab Hamidi-Rad, Aditya Sant\\  
InterDigital Communications\\
\texttt{\{firstname.lastname\}@interdigital.com}
\AND
David J. Love, Christopher G. Brinton  \\
Elmore School of Electrical and Computer Engineering, Purdue University\\
\texttt{\{cgb,djlove\}@purdue.edu}
}

\iclrfinalcopy 
\begin{document}

\maketitle

\begin{abstract}
    In recent years, machine learning (ML) methods have become increasingly popular in wireless communication systems for {several applications.} 
    A critical bottleneck for designing ML systems for wireless communications is the availability of realistic wireless channel datasets, which are extremely resource intensive to produce. To this end, the generation of realistic wireless channels plays a key role in the subsequent design of effective ML algorithms for wireless communication systems.
    Generative models have been proposed to synthesize channel matrices, but outputs produced by such methods may not correspond to geometrically viable channels and do not provide any insight into the scenario of interest. 
    In this work, we aim to address both these issues by integrating a parametric, physics-based geometric channel (PBGC) modeling framework with generative methods.
    To address limitations with gradient flow through the PBGC model, a linearized reformulation is presented, which ensures smooth gradient flow during generative model training, while also capturing insights about the underlying physical environment.
    We evaluate our model against prior baselines by comparing the generated samples in terms of the 2-Wasserstein distance and through the utility of generated data when used for downstream compression tasks. 
\end{abstract}

\section{Introduction}\label{sec:intro}
The use of Machine learning (ML)  for applications in wireless communication has seen extensive interest in the past few years. At the physical layer (PHY) of wireless systems, ML research primarily aims to estimate and mitigate distortions in electromagnetic signals during over-the-air (OTA) transmission \cite{ch_compression,ch_estimation,dl_wireless_survey,sant2022deep} and address noise and non-linearities at transmit or receive antennas \cite{estimation_survey,sant2024insights}. However, for practical deployments, the ML pipeline requires a substantial amount of OTA wireless channel data which is a complex, costly, and time-intensive \cite{OTAdata_capture1,OTAdata_capture2,OTAdata_capture3}. 

Generative models have been proposed to mitigate this problem by artificially synthesizing large wireless datasets using considerably fewer OTA data samples \cite{channelgan}. However, unlike common modalities of data (image, text, audio, etc.) which are directly human-interpretable, the wireless channel data is a tensor of complex numbers and is \emph{not human-interpretable or easily visualized}. This poses two major challenges around effectively testing and using generative channel models.  
Firstly, the outputs of generative models may not correspond to valid channels.  Here, the validity of channel data implies that the wireless channel can be represented as a multipath geometric model representing the multiple paths the transmitted signal takes, before arriving at the receiver \cite{chanmodel_1,chanmodel_2}. Secondly, it is hard to gain any insights about the physical parameters associated with the signal propagation or any information about the environment or scenario being considered (e.g. Angles associated with paths, gains of paths, line-of-sight transmission or non-line of sight, etc.) from generated data samples.

This work primarily focuses on the design of generative models for synthesizing millimeter wave (mmWave) channels, crucial for next-generation wireless communication and IoT \cite{iot_mmwave1}. 
The proposed method overcomes the limitations of existing approaches by incorporating a physics-based generative channel (PBGC) model into the generative pipeline. As the PBGC model parametrizes the channel generation, our generative process will learn the joint distribution of the underlying parameters responsible for channel generation. We also propose a linearized reformulation of the PBGC model to address training challenges due to its non-convex loss landscape.

\section{System Model and Approach}\label{sec:model}

We consider a wireless communication system with $N_t$ transmit and $N_r$ receive antennas. The associated PBGC model defined by $M: \mathbb{R}^{3P} \rightarrow \mathbb{R}^{2 \times N_t \times N_r}$ \cite{pb_model}, maps a set of parameters $\textbf{s} \in \mathbb{R}^{3P}$ to a matrix $\textbf{H} \in \mathbb{C}^{N_t \times N_r}$, where $P$ is the number of paths that a transmitted signal takes before being received at the receiver antennas. The PBGC model is given by

\begin{equation}
\label{eq:channel_sum}
    \textbf{H} = M(\textbf{s}) = \sum_{p=1}^{P} g_p \textbf{a}_r(\theta^p_a) \textbf{a}_t(\theta^p_d)^H.
\end{equation}

Where, $\textbf{s} = [g_p,\theta_a^p,\theta_d^p]_{p = 1}^P$, and
$g_p$ represents the propagation gain associated with the $p$-th path, $\textbf{a}_t(\cdot), \textbf{a}_r(\cdot)$ represent the array response vectors on the transmit and receive antennas, $\theta_d^p$ and $\theta_a^p$ represent the corresponding angle of departure and angle of arrival. More details on the PBGC model can be found in Appendix \ref{app:Channel_model}. 

\begin{figure}[t]
    \centering
    \includegraphics[width=0.7\columnwidth]{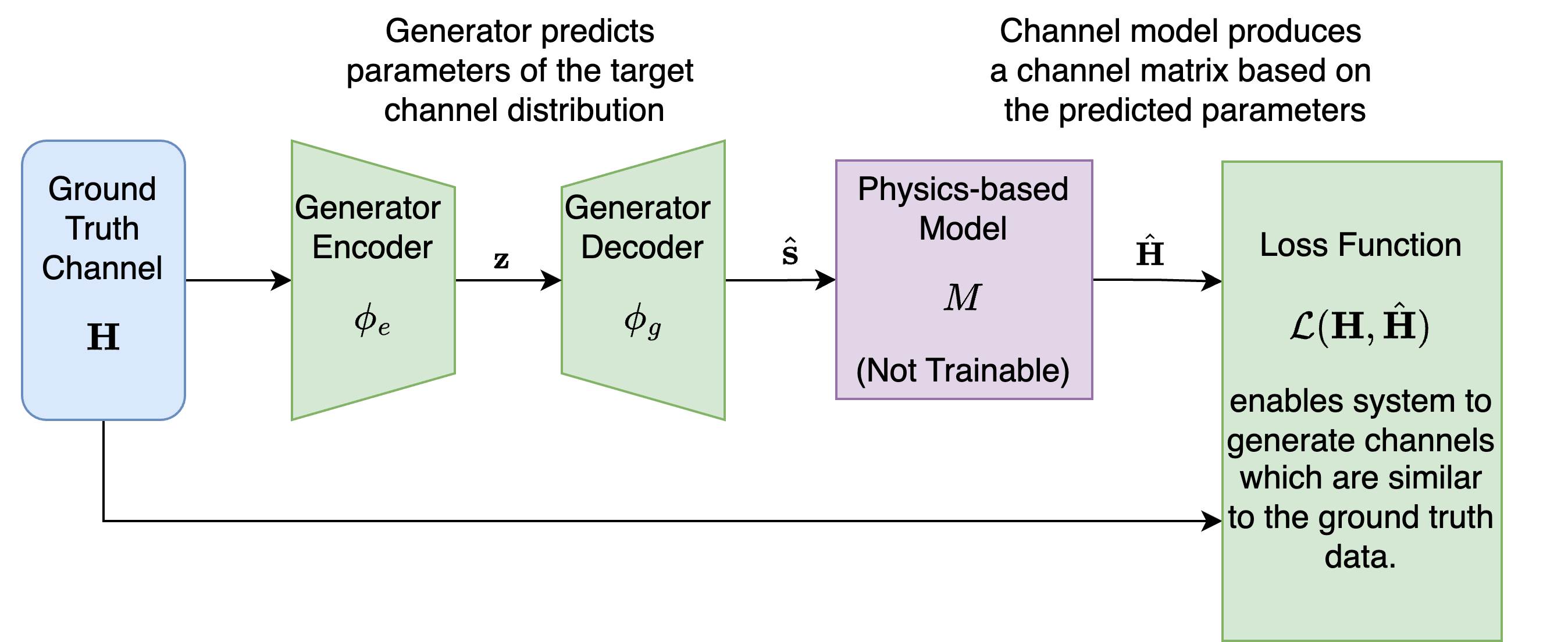}
    \caption{In the straightforward implementation, the generator directly predicts the parameters $\hat{\textbf{s}}$, which are then used by the PBGC  model $M$ to produce the predicted channel (Top).}
    \label{fig:model_train}
\end{figure}

\subsection{Generative Model to Predict Channel Statistics} \label{sec:pred_params}

For the generative model we use the variational autoencoder (VAE) architecture \cite{vae}, as seen in Fig. \ref{fig:model_train}. We use the generative model to produce the parameter vector $\hat{\textbf{s}}$ using a latent variable $\textbf{z}$, which is then passed to the PBGC model $M$ to produce a valid channel matrix. More details on the training can be found in Appendix \ref{app:training}.

\begin{figure*}
    \centering
    \includegraphics[width=0.3\textwidth]{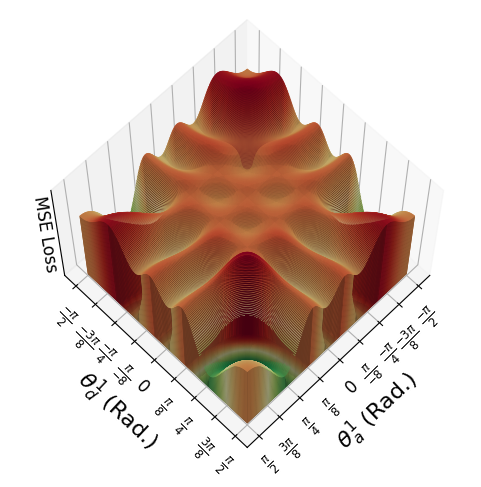}
    \includegraphics[width=0.3\textwidth]{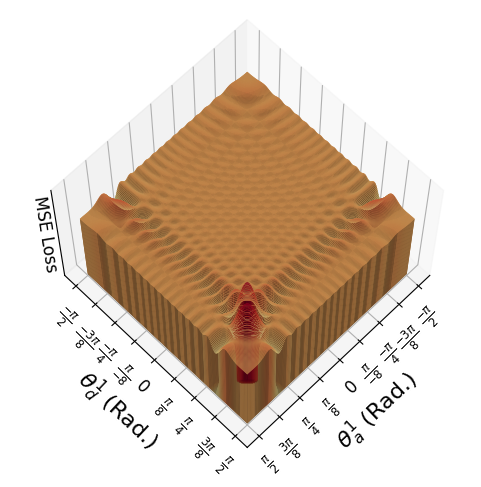}
    \includegraphics[width=0.3\textwidth]{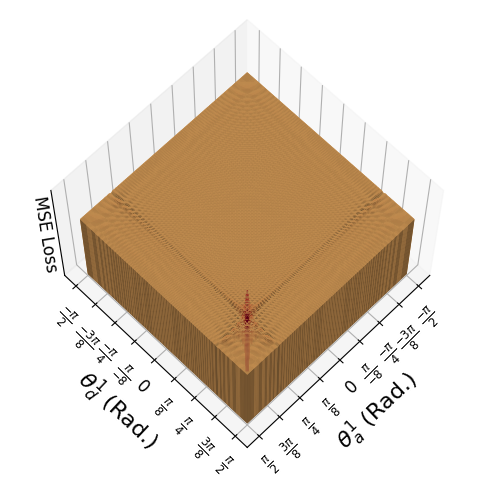}
    
    \includegraphics[width=0.3\textwidth]{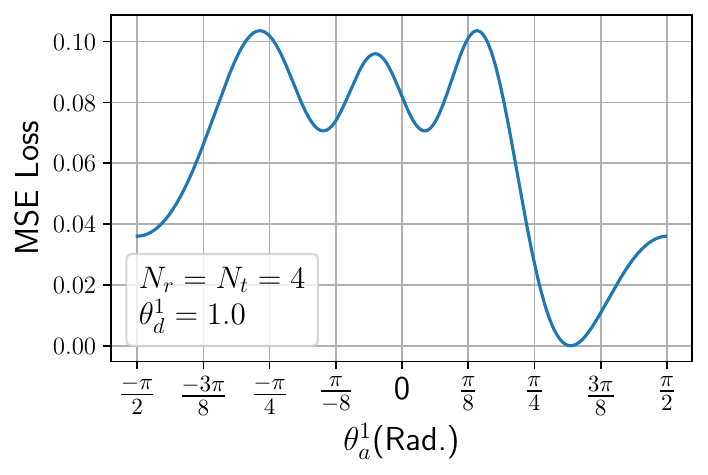}
    \includegraphics[width=0.3\textwidth]{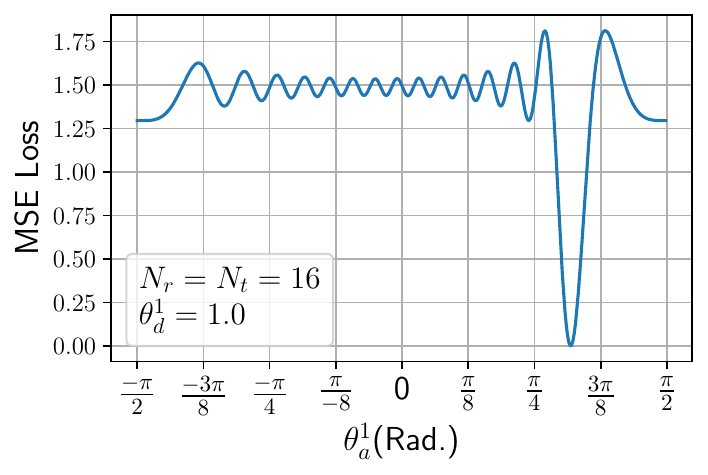}
    \includegraphics[width=0.3\textwidth]{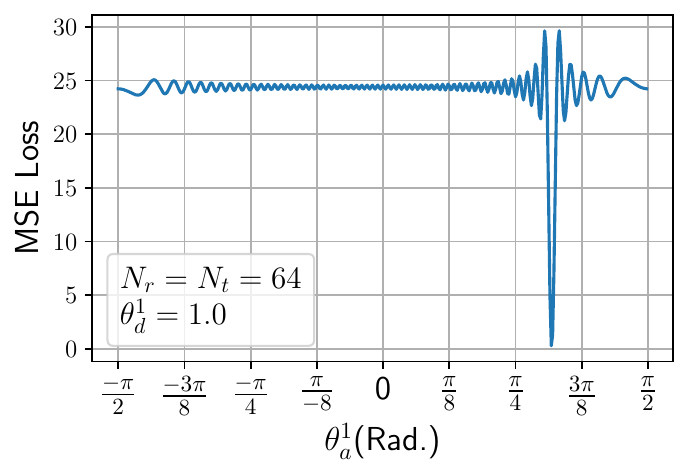}
    \caption{The loss surface as a function of $(\theta^a_1,\theta^d_1)$ in reference to a channel matrix with $\theta^a_1 = \theta^d_1 = 1.0$ using a PBGC model $M$ with $P=1$ and $N_r = N_t = 4,16,64$ antennas respectively. The PBGC model $M$ is extremely non-convex as a function of the parameters $\theta_a,\theta_d$ because of periodicity arising from the formulation of the array response vectors. As the number of antennas $N_r,N_t$ increase, gradient flowing through the model will be negligible when at locations further away from the minima, where the loss surface is effectively flat. }
    \label{fig:nonconvex_surface}
\end{figure*}

\subsubsection{Limitations of generative model training using PBGC}
The direct use of the PBGC model with the VAE leads to poor training performance due to the presence of sinusoidal functions in the channel model (Appendix \ref{app:Channel_model}).
This property is illustrated in Fig. \ref{fig:nonconvex_surface}. The non-convexity arising from this periodicity is only exacerbated with the addition of multiple paths. However, these periodicities cannot be effectively approximated by the non-linear activation functions used in deep neural networks, leading to difficulties in training \cite{rectified_boltzman,guide_to_conv_aritmetic}.
As a result, depending on the location of the optimizer in the parameter space, the gradient may not flow during backpropagation, resulting in the optimizer converging at non-optimal points.
Further, as the number of antennas, $N_t, N_r$ increase, the number of local minimas grows, and the optimality gap, the difference between the loss values at the local minima and the global minima, widens, significantly impacting the overall loss (Appendix \ref{app:convergence}). 


\subsection{Linearized Reformulation of the Physics Model}

\label{sec:pred_matrix}
To overcome the challenges posed by the PBGC model, while maintaining the key underlying model features for channel generation, the channel synthesis is presented via a weighted sum over a parameter dictionary.
In order to compute the dictionary elements, the range of angles $\theta_a^p,\theta_d^p$, given by $[\theta_{\min},\theta_{\max}]$, is equally divided into $R$ intervals of width $\Delta\theta = (\theta_{\max}-\theta_{\min})/R$. We pre-compute the outer product between the array response vectors, $\textbf{a}_r(\theta^p_a) \textbf{a}_t(\theta^p_d)^H$, at the discretized angle values and store them in a dictionary $\textbf{D}$.
The  dictionary $\textbf{D}$ has a total of $R^{2}$ elements, where each element of the dictionary, $\textbf{D}_{i,j}~ \forall~ i,j \in \{1,R\}$, is given by:

\begin{equation}
\label{eq:calc_array_dict}
    \textbf{D}_{i,j} = \textbf{a}_r(\theta_i) \textbf{a}_t(\theta_j).
\end{equation} 

The relaxed PBGC model using the linearized reformulation can now be expressed as,

\begin{equation}
\label{eq:lincombi_channel}
    \textbf{H} = \sum_{i=1}^R\sum_{j=1}^R \textbf{W}_{i,j}\textbf{D}_{i,j},
\end{equation}
where, the channel generation is parametrized by the gain matrix $\textbf{W} \in \mathbb{R}^{R^2}$, instead of the parameters $\textbf{s}$, and the channel is constructed by the element-wise product between $\textbf{W}$ and $\textbf{D}$. 
{By making these changes in the pipeline, we now model the output channel $\textbf{H}$ as a linear function of the gain matrix $\textbf{W}$, mitigating the issues arising from the non-convexity of the PBGC model $M$. Further, it must be noted that the accuracy of the relaxed model is strongly tied to the number of intervals, $R$. For a suitably high value of $R$, any channel $\textbf{H}$ can be approximated by (\ref{eq:lincombi_channel}). 


\subsection{Generative Model to Predict the Gain Matrix}

\begin{figure}
    \centering
    \includegraphics[width=0.66\columnwidth]{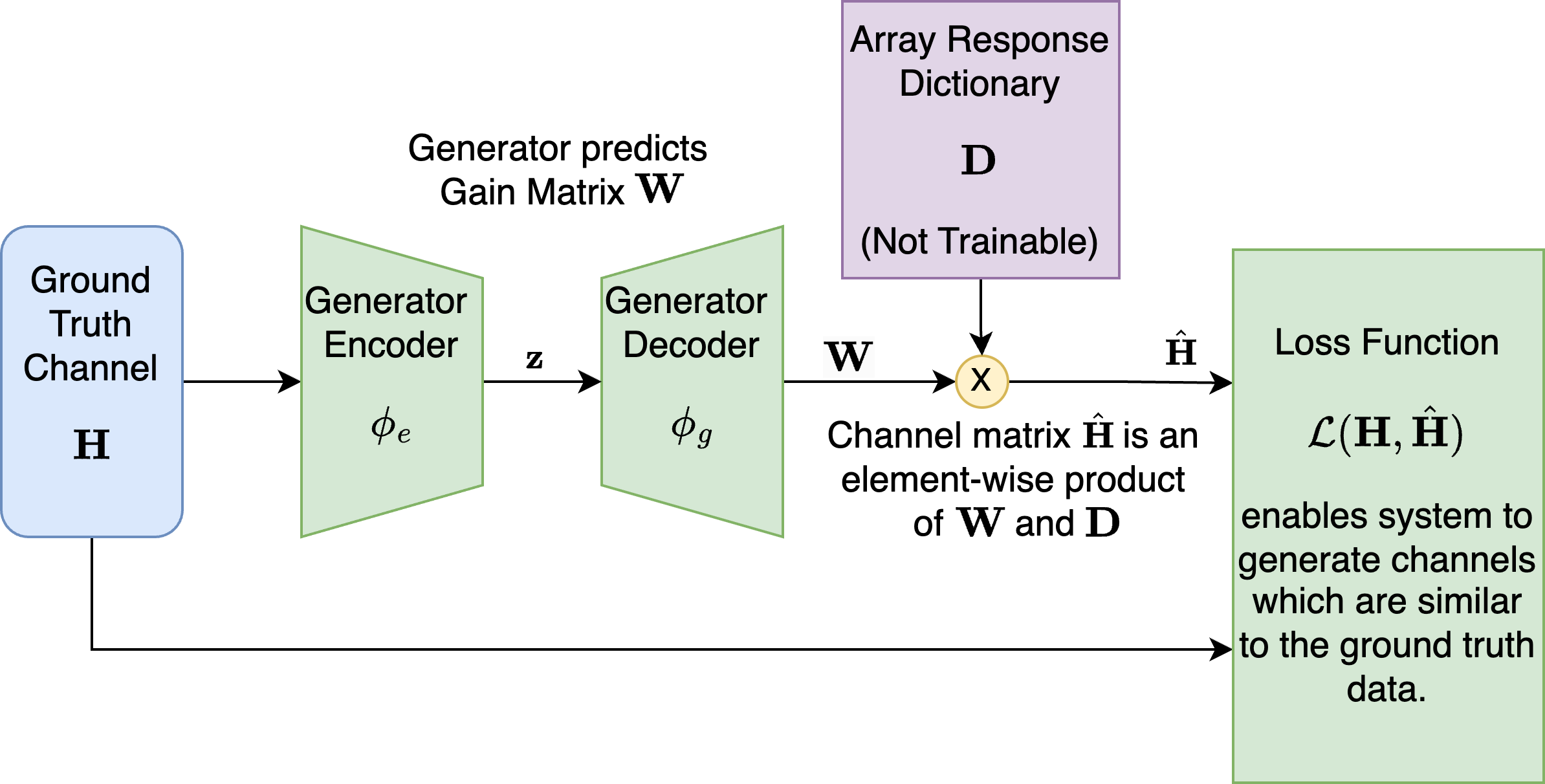}
    \caption{We relax the PBGC model by defining a discretized array response dictionary $\textbf{D}$ and using the generator to output gain the gain matrix $\textbf{W}$. The elementwise multiplication of $\textbf{W}$ and $\textbf{D}$ mimics the PBGC model process. This relaxation allows the flow of gradient through the generator, enabling it to converge to more suitable optima.}
    \label{fig:limmodel_train}
\end{figure}

The modified generative model using the linearized reformulation is shown in Fig. \ref{fig:limmodel_train}. The VAE decoder $g_{\phi_d}:\mathbb{R}^{Z} \rightarrow \mathbb{R}^{R \times R}$ now predicts the gain matrix $\textbf{W}$, where $R$ is the resolution of $\textbf{W}$. During training, the decoder $g_{\phi_d}$ takes in the latent vector $\textbf{z}$ as input and produces a gain matrix $\textbf{W}$ as $\textbf{W} = g_{\phi_d}(\textbf{z})$.
The gain matrix is then used to generate the predicted channel using \eqref{eq:lincombi_channel}.

\section{Experimental Results}\label{sec:results}
In this section, we analyze the performance of our method on wireless datasets generated corresponding to real-life scenarios and compare our method against prior art baselines. For our PBGC model \eqref{eq:channel_sum}, we consider transmit and receive antennas $N_t = N_r = 16$ and assume the number of paths $P = 5$. We consider five datasets, sythesized using the DeepMIMO framework \cite{deepmimo}, modeling street intersections in different cities (see Appendix \ref{Sec:appendix_dataset_info}). 
We use a VAE as the generative model, $(g_{\phi_e},g_{\phi_d})$, which is trained using the Adam optimizer with a learning rate of $1e^{-3}$. {Model architecture details are described in Appendix \ref{app:training}} We compare our model against  ChannelGAN (CGAN)\cite{channelgan}, the DUNet diffusion model (DUNet) \cite{diffusion_models_for_channels} and a VAE version of CSINet \cite{csinet}. Models are trained for $300$ epochs with a batch size of $256$. For our model, we use resolution $R = 64$ and latent dimension $z = 64$.

In Table \ref{tab:channel_dist}, we analyze the ability of our method to capture the distribution of channel matrices compared to baseline methods. We generate $3000$ synthetic channel matrices and compare the 2-Wasserstein distance \cite{2wasserstein} and Maximum Mean Discrepancy \cite{mmd} between the distribution of the generated channels and the true channels.
The channels generated by our method are closer to the distribution of true channels than those generated by the ChannelGAN baseline by up to $4 \times$. This shows that our method can generate more realistic channel data as compared to baselines. A comprehensive evaluation of the generative modeling schemes on the downstream task of channel compression (CSI compression) can be found in Appendix \ref{app:experiments}. 

\begingroup
\setlength{\tabcolsep}{1.2pt} 
\renewcommand{\arraystretch}{1.1} 
\begin{table}[t]
    \small
    \centering
    \begin{tabular}{|c||c|c|c|c|c|c|c|c|}
    \hline
       & \multicolumn{4}{|c|}{2-Wasserstein Distance}  & \multicolumn{4}{|c|}{MMD}\\
      \hline
      Dataset   & Ours & CGAN & DUNet & CVAE & Ours & CGAN & DUNet & CVAE\\
      \hline
      Boston  & \textbf{0.475} & 0.84 & 0.687 & 0.87 & \textbf{0.013} & 0.045 & 0.052 & 0.095 \\
       ASU  & \textbf{0.283} & 0.932 & 0.641 & 1.095 & \textbf{0.012} & 0.169 & 0.02 & 0.592 \\
       Indoor & \textbf{0.17} & 0.339 & 0.529 & 1.802 & \textbf{0.005} & 0.003 & 0.02 & 0.375 \\
       BS10  & \textbf{0.656} & 1.632 & 0.857 & 1.837 & \textbf{0.072} & 0.317 & 0.108& 0.414\\
       BS11  & \textbf{0.267} & 0.882 & 0.463 & 1.12 &\textbf{0.016} & 0.086 & 0.033& 0.132\\
       \hline
    \end{tabular}
    \caption{The distribution of channels modelled by our method is more similar to the real distribution compared to baselines in terms of 2-Wasserstein distance and Maximum Mean Discrepancy (MMD).}
    \label{tab:channel_dist}
    \vspace{-3mm}
\end{table}
\endgroup



\section{Conclusion}\label{sec:conclusion}
In this paper, we developed a generative pipeline that leverages a PBGC model for parametrized channel generation. We tackle the extreme non-convexity in the PBGC model by developing a dictionary-based relaxation of the PBGC model and learning a sparse gain matrix whose non-zero values denote the parameters of the associated paths.
We empirically show that our method effectively captures path-specific parameter distributions for a given dataset of channel matrices and outperforms prior arts in terms of 2-Wasserstein distance and MMD.
Our work can be extended to 3-dimensional scenarios with angles of elevation and additional parameters such as path delay. 

\section{Acknowledgements}

This project was undertaken as part of an internship at Interdigital, Inc.   

Prof. Christopher G. Brinton's contributions to this project were supported in part by the Defense Advanced Research Projects Agency (DARPA) under grant D22AP00168, the Office of Naval Research (ONR) under grant N00014-21-1-2472, and the National Science Foundation (NSF) under grant CNS-2212565

\bibliographystyle{named}
\bibliography{iclr2025_delta}

\appendix

\section{Related Work}\label{sec:rel_work}

Several works propose using a generative model to produce novel channel samples through the stochastic generative process. A generative  adversarial network (GAN) based wireless channel modeling framework was first introduced in \cite{GAN1}. Authors of \cite{channelgan} utilize a Wasserstein-GAN with Gradient Penalty (WGAN-GP) to synthesize novel channel matrices given a limited set of training data points. \cite{mimogan} trained their model on multiple-input multiple-output (MIMO) data, with a discriminator explicitly designed to learn the spatial correlation across the channel data.
In \cite{diffusion_models_for_channels}, the authors utilize diffusion models as the generative backbone to circumvent the issue of mode collapse in GANs. Works such as \cite{score_based} utilize a score-based generative model for joint channel modeling and estimation. All of the above works utilize a generative model to directly produce channel matrices at the output have no guarantees on the validity of the generated channel data, with limited interpretability. 

An alternative research direction involves using labeled datasets to predict parameters associated with the wireless channel. In \cite{gen_models_mmwave_uav} and \cite{multi_freq_model}, the authors use a conditional variational autoencoder (VAE) and a conditional GAN framework, respectively, to generate the channel parameters given the location of UAVs. 
The above works require datasets labeled with metadata relating to the environment and locations of the transmitter or receiver to produce parameters that can be used in downstream channel models and do not evaluate the channel matrices directly. In contrast, our method does not require labeled data, and can learn channel parameters directly from the channel matrix.

\section{Details of the Physics Based Geometric Channel Model (PBGC)} \label{app:Channel_model}

As mentioned, we consider a communication setup with $N_t$ transmit and $N_r$ receive antennas. The PBGC model considers the channel matrix $\textbf{H}$ to be a superposition of the individual propagation matrices associated with each of the $P$ paths. The overall model is thus expressed as:

\begin{equation}
    \textbf{H} = M(\textbf{s}) = \sum_{p=1}^{P} g_p \textbf{a}_r(\theta^p_a) \textbf{a}_t(\theta^p_d)^H.
\end{equation}

Where, $\textbf{s} = [g_p,\theta_a^p,\theta_d^p]_{p = 1}^P$, and
$g_p$ represents the propagation gain associated with the $p$-th path, $\textbf{a}_t(\cdot), \textbf{a}_r(\cdot)$ represent the array response vectors on the transmit and receive antennas, $\theta_d^p$ and $\theta_a^p$ represent the corresponding angle of departure and angle of arrival, {both of which take values between $[-\pi,\pi]$ radians}. Also, in mmWave channels, the total number of paths $P$ is typically small in over-the-air transmission. 

To simplify the discussion, we consider Uniform Linear Arrays (ULA) \cite{chanmodel_1} and limit the discussion to the azimuth plane. Thus, the array response vectors can be defined as:
\begin{align}
\label{eq:array_responses}
    \textbf{a}_t(\theta_d^p) = \frac{1}{\sqrt{N_t}}[1,e^{ju\sin(\theta_d^p)},\dots,e^{j(N_t-1)u\sin(\theta_d^p)}]^T,\\
    \textbf{a}_r(\theta_a^p) = \frac{1}{\sqrt{N_r}}[1,e^{ju\sin(\theta_a^p)},\dots,e^{j(N_r-1)u\sin(\theta_a^p)}]^T,
\end{align}

where, $u = \frac{2\pi}{\lambda}d$, $\lambda$ is the wavelength of the carrier signal and $d$ is the distance between antenna elements. 

\section{Generative Model Training with PBGC}\label{app:training}
During the training phase, the encoder of the generative model takes a channel matrix \textbf{H} as input and samples a latent vector $\textbf{z}$ from the posterior distribution as $\textbf{z} \sim f_{\phi_e}(\textbf{H}).$ The decoder of the generative model then takes in the latent vector $\textbf{z}$ as input and produces a parameter vector $\hat{\textbf{s}}$ as $\hat{\textbf{s}} = g_{\phi_d}(\textbf{z})$. The predicted parameter vector is then passed to the model $M$ to produce an output channel $\hat{\textbf{H}}$ as follows $\hat{\textbf{H}} = M(\hat{\textbf{s}}).$ The system loss is a generalization of the evidence based lower bound (ELBO)\cite{vae}, given by
\begin{equation}
\label{eq:vae_loss_direct}
    \mathcal{L} = ||\textbf{H}-\hat{\textbf{H}}||_2^2 + \alpha_{D} \cdot \textsf{KL}(\textbf{z},\mathcal{N}(0,\textbf{I})).
\end{equation}
Here, the first term corresponds to the reconstruction or mean square error (MSE) loss between the input $\textbf{H}$ and the predicted channel matrix $\hat{\textbf{H}}$. This ensures that the outputs are similar to the inputs. The second term penalizes the Kulback-Leibler (KL) divergence \cite{kldiv} between the latent vector $\textbf{z}$ and a simple, known distribution, in this case, the multivariate unit Gaussian distribution $\mathcal{N}(0,\textbf{I})$, where $\textbf{I}$ is the identity matrix of dimension $Z$. This encourages the distribution of the latent vectors to be similar to $\mathcal{N}(0,\textbf{I})$.

\subsection{Convergence Issues with the PBGC Model}\label{app:convergence}}

\begin{figure}[h!]
    \centering
    \includegraphics[width=0.5\columnwidth]{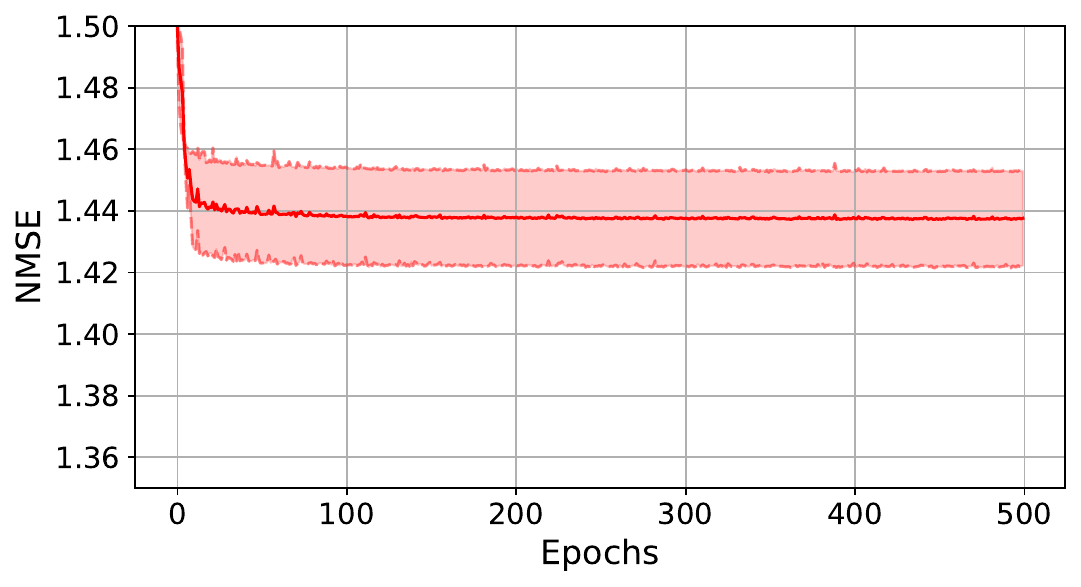}
    \caption{In the straightforward integration of the PBGC model with the VAE, the generator directly predicts the parameters $\hat{\textbf{s}}$, which are then used by the PBGC  model $M$ to produce the predicted channel. In such implementations, the generator cannot converge to suitable optima due to the non-convexity of the PBGC model. Further, as the number of antennas, $N_t, N_r$ increases, the optimality gap, the difference between the loss values at the local minima and the global minima, widens, significantly impacting the overall loss. In this figure, we show the average training performance of the pipeline over several independent training instances. As can be seen, the normalized mean squared error (NMSE) barely reduces and continues to remain high across the entire training horizon.}
    \label{fig:model_train_loss}
\end{figure}

\subsection{Model Architecture}

\begin{figure*}[h!]
    \centering
    \includegraphics[width=0.95\textwidth]{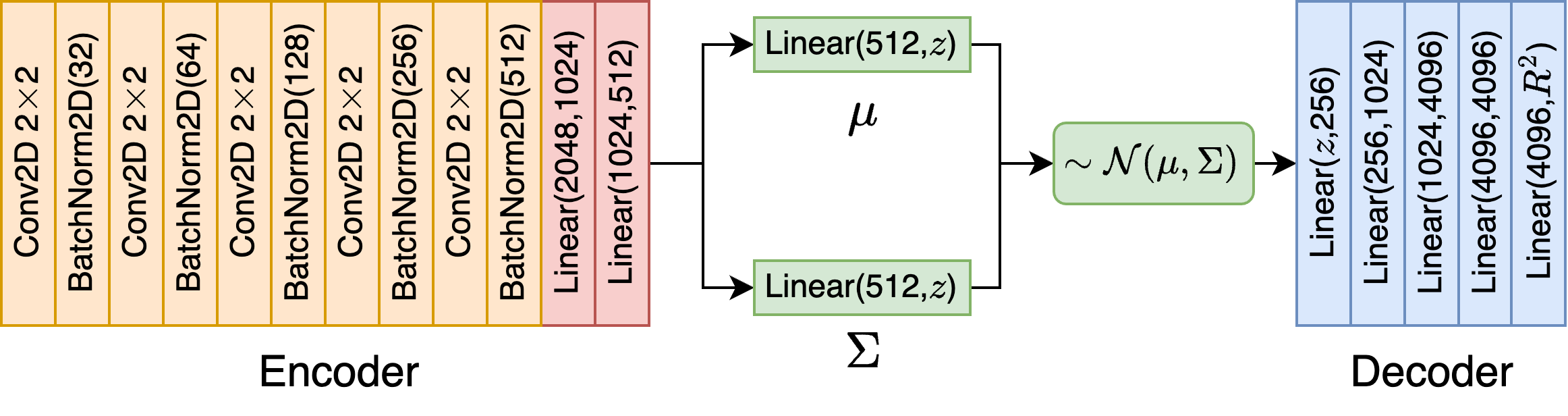}
    \caption{The architecture of the VAE model used in our experiments. We use the Leaky ReLU activation after each \textsf{BatchNorm} and \textsf{Linear} layer.}
    \label{fig:model_arch}
\end{figure*}

We consider a variational autoencoder (VAE) architecture as the generative model for our experiments. The architecture of our model is given in Fig. \ref{fig:model_arch}.

\section{Experimental Results} \label{app:experiments}
\subsection{Dataset information}\label{Sec:appendix_dataset_info}
Five datasets are generated using the  used to generate channel datasets using 3D ray tracing. Datasets corresponding to the following scenarios are used; (i) Two base stations in an outdoor intersection of two streets with blocking and reflecting surfaces, given by (BS10) and (BS11); (ii) an indoor conference room, given by (Indoor); (iii) a section of downtown Boston, Massachusetts, USA, generated using the 5G model developed by RemCom \cite{remcom}, given by (Boston) and (iv) a section of the Arizona State University campus in Tempe, Arizona, USA, given by (ASU).

We evaluate the accuracy of generated channels as well as effectiveness of the generative pipeline for downstream channel compression tasks.

\subsection{Prediction of Parameters} 

\begin{figure}[!h]
    \centering
    \includegraphics[width=0.36\columnwidth]{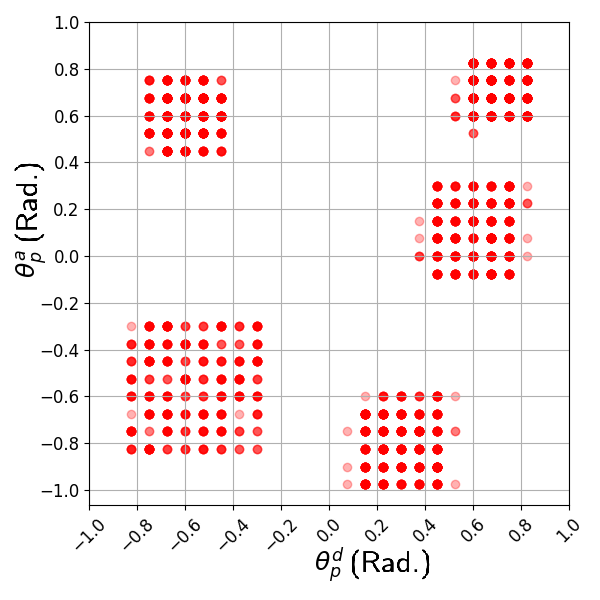}
    \centering
    \includegraphics[width=0.36\columnwidth]{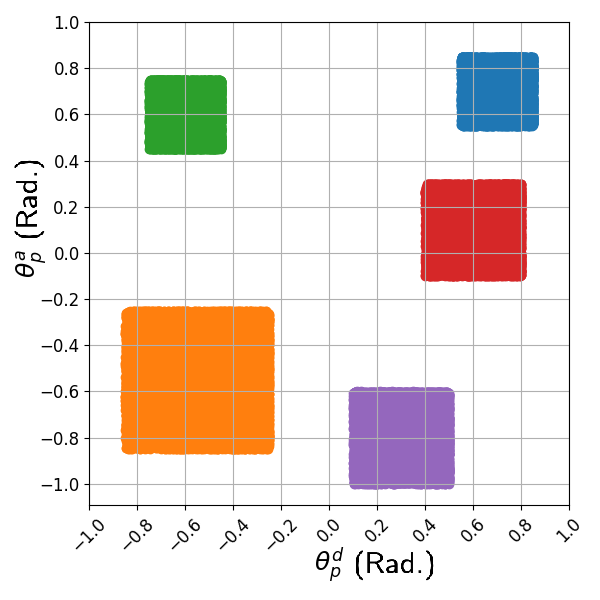}
    \label{fig:enter-label}
    \caption{The distributions of angles of arrival and departure $(\theta^a_p,\theta^d_p)$ captured by our method (Left) match the underlying distributions of the training dataset (Right). Each color corresponds to a distinct path.}
    \label{fig:param_comp}
\end{figure}
We evaluate the accuracy of our system in capturing the underlying distribution of parameters associated with a given dataset. A user-defined  dataset $\textbf{D}$ of channel matrices is created by sampling parameters $[g_p,\theta_p^a,\theta_p^d]_{p=1}^P$ from a user-defined distribution. The proposed model is trained on this dataset, and the distribution of the generated parameters from this model is illustrated. The results Fig. \ref{fig:param_comp}, compare the ground truth distributions with the output of the generative model. We observe that our method can accurately capture the distributions of the angles of arrival and departure for each path. This shows that our model training and parameter extraction methods can be used to determine the distributions of parameters of input channels without requiring labeled data.

\begin{figure*}
    \centering
    \begin{subfigure}{0.32\textwidth}
        \centering
        \includegraphics[height=1.45in]{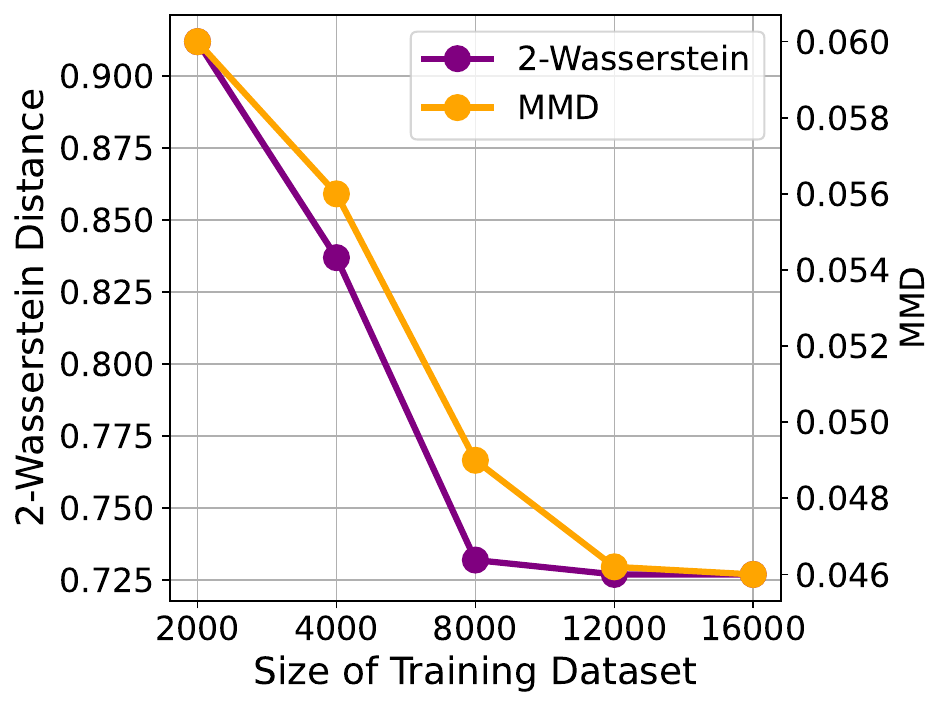}
        \caption{}
        \label{fig:abl_num_dp}
    \end{subfigure}
    \begin{subfigure}{0.32\textwidth}
        \centering
        \includegraphics[height=1.45in]{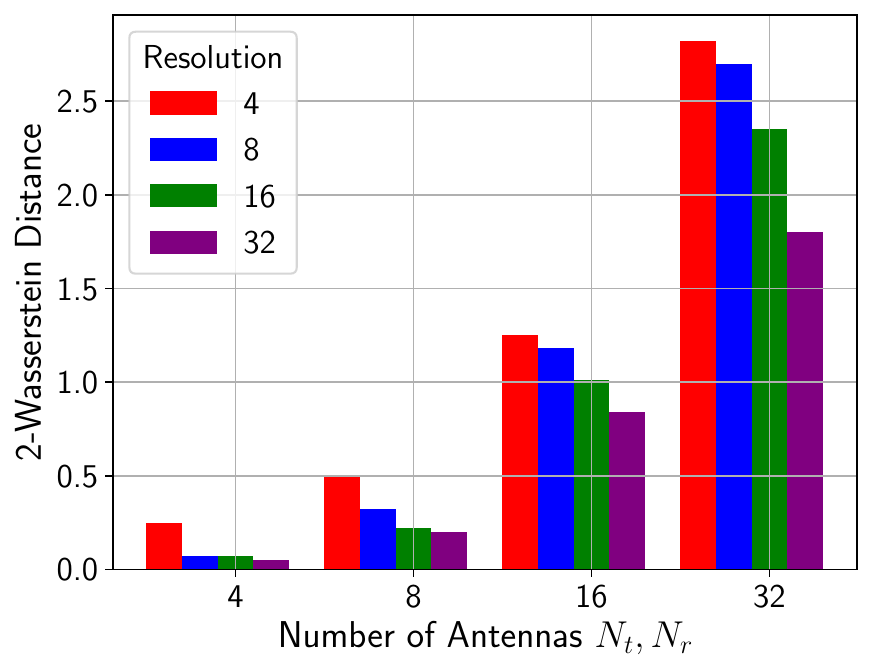}
        \caption{}
        \label{fig:abl_resolution}
    \end{subfigure}
    \begin{subfigure}{0.32\textwidth}
        \centering
        \includegraphics[height=1.45in]{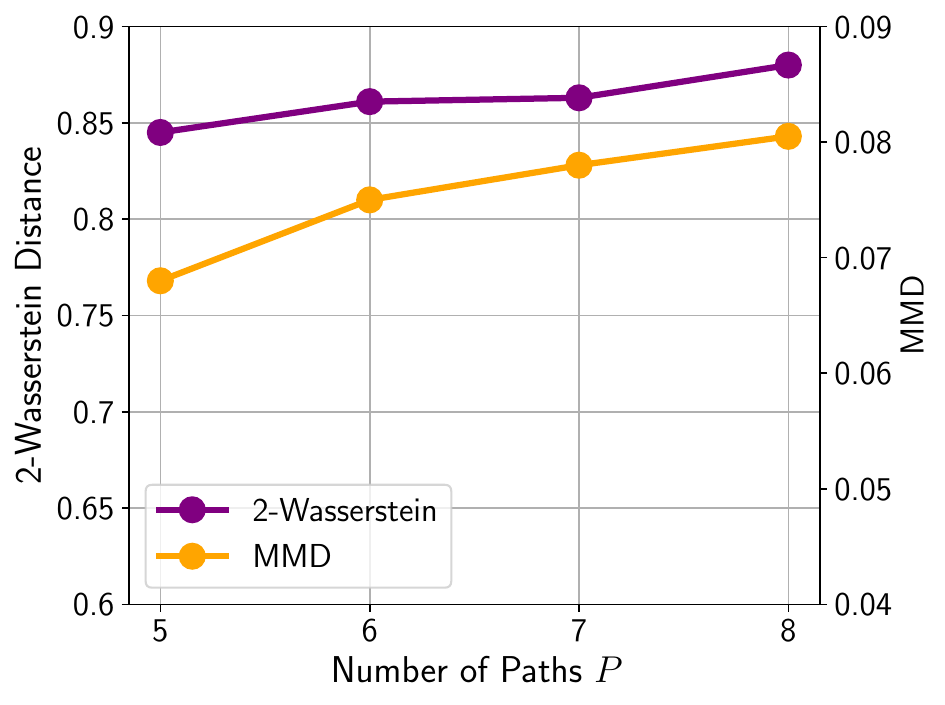}
        \caption{}
        \label{fig:abl_num_paths}
    \end{subfigure}
    \caption{(a) Our method can generate samples with a high degree of fidelity in terms of 2-Wasserstein distance and MMD. even with a training dataset of around $50\%$ of the size of the training dataset. (b) For a physics model $M$, as the number of antennas $(N_r,N_t)$ increase, an increase in the resolution ($R$) of the gain matrix $\textbf{W}$ results in a higher degree of fidelity with the input data distribution in terms of the 2-Wasserstein distance, as the highest possible precision with which parameters can be estimated is dependent on the number of antennas. (c) The performance of our method is consistent across a varying number of paths $P$, as the generative process is independent of $P$, relying on the loss function to balance reconstruction fidelity and the identified number of paths, given by the number of non-zero values in the gain matrix \textbf{W}.}
\end{figure*}

\subsection{Effect of Varying Size of Training Dataset}
In this experiment, we train our model on datasets of varying sizes, choosing a subset of the original dataset of the appropriate size. We use the DeepMIMO dataset BS10 for this experiment, which consists of $16,000$ datapoints.

In Fig. \ref{fig:abl_num_dp}, we observe that our method is able to provide a similar level performance even when the size of the training data is reduced by up to $\sim 40\%$. This is because our model predicts distributions in the parameter space, which are less complex than the distributions in the channel space, even a small number of datapoints can capture the distribution related characteristics of the input channels. In a practical deployment, this translates to significant savings in terms of the resources deployed to acquire channel data.

\subsection{Effect of Increasing Number of Paths}
In this experiment, we observe the effect of increasing the number of constituent paths $P$ in a channel matrix. We consider the channel dataset given in Table \ref{tab:channel_dist} and add additional paths where needed as follows. Path $6$ is sampled from $\theta_p^a \sim \mathcal{U}(0.4,0.8)/ \theta_p^a \sim \mathcal{U}(0.1,0.3)$, Path 7 is sampled from $\theta_p^a \sim \mathcal{U}(0.6,1.0)/ \theta_p^a \sim \mathcal{U}(-0.3,-0.1)$, and Path $8$ is sampled from $\theta_p^a \sim \mathcal{U}(-0.3,0.9)/ \theta_p^a \sim \mathcal{U}(0.6,1.0)$. $g_p \sim \mathcal{U}(0.001,0.01)$ for all paths.

In Fig. \ref{fig:abl_num_paths}, we observe that the performance of our generative pipeline remains consistent across a varying number of paths $P$. This is because our model is independent of $P$, and leverages the formulation of the loss function to balance the reconstruction accuracy and the number of non-zero output values, which dictates the number of identified paths. This is encapsulated by the last term, which enforces output sparsity. The hyperparameter $\alpha_S$ in the loss function can thereby be tuned by observing the reconstruction loss. Thus, our method can adapt to a range of values for the number of paths by finding a suitable balance between the NMSE and the L1 regularization loss such that the number of non-zero values are proportional to the number of paths.

\subsection{Cross Evaluation Between Distinct Datasets}

In this experiment, we analyze the ability of our method to learn distinct channel distributions based on the cross evaluation of models trained on distinct channel distributions in the context of a downstream channel compression task. 

We consider two channel datasets $R_{10},R_{11}$ from the DeepMIMO scenario, generated from base stations $10$ and $11$ respectively. We train an independent instance of the generative model on each dataset and generate synthetic datasets of size $20,000$, given by $G_{10},G_{11}$ respectively. We then train independent instances of the CSINet channel compression model \cite{csinet} on each set.  We perform cross evaluation considering all pairwise combinations of training and testing datasets and calculate the test NMSE given in Table \ref{tab:cross_eval}.

Now, a model trained on $G_{10}$ should generalize well to $R_{10}$, and vice versa for a model trained on $G_{11}$. In Table \ref{tab:cross_eval}, we observe that a compression model trained on data generated by our model follows the aforementioned rules, indicating that our method can capture the distinctions between two different datasets and generate distinct channel data samples.  

\begingroup
\setlength{\tabcolsep}{3pt} 
\renewcommand{\arraystretch}{1.1} 
\begin{table*}[t]
    \scriptsize
    \centering
    \begin{tabular}{|c||c c c c|c c c c|c c c c| c c c c|}
    \hline
       \multirow{2}{2em}{Train} & \multicolumn{4}{|c|}{Testing $R_{10}$} & \multicolumn{4}{|c|}{Testing $R_{11}$} & \multicolumn{4}{|c|}{Testing $G_{10}$} & \multicolumn{4}{|c|}{Testing $G_{11}$} \\
        \cline{2-17}
    & Ours & CGAN & DUNet & CVAE & Ours & CGAN & DUNet & CVAE & Ours & CGAN & DUNet & CVAE & Ours & CGAN & DUNet & CVAE  \\
    \hline
    
    $R_{10}$ & 0.02 & 0.02 & 0.02 &0.02& \lightgray{1.35} & \lightgray{1.35} & \lightgray{1.35} & \lightgray{1.35} & \textbf{0.06} & 1.29 & 0.46 &0.55& \lightgray{1.36} & \lightgray{1.37} & \lightgray{1.18} & \lightgray{1.07}\\
    \hline
    
    $R_{11}$ & \lightgray{1.14} & \lightgray{1.14} & \lightgray{1.14} & \lightgray{1.14} & 0.06 & 0.06 & 0.06 &0.06& \lightgray{1.21} & \lightgray{1.51} & \lightgray{1.14} & \lightgray{1.04}& \textbf{0.25} & 0.45 &  0.42 & 0.72\\
    \hline
    
    $G_{10}$ & \textbf{0.05} & 0.77 & 0.19 &0.85& \lightgray{1.41} & \lightgray{0.97} & \lightgray{1.04} & \lightgray{1.49} & 0.01 & 0.09 & 0.03 &0.09& \lightgray{1.5} & \lightgray{0.92} & \lightgray{1.17} & \lightgray{1.1}\\
    \hline
    $G_{11}$ & \lightgray{1.1} & \lightgray{1.37} & \lightgray{1.02} & \lightgray{1.04} & \textbf{0.14} & 0.56 & 0.37 &0.52& \lightgray{1.31} & \lightgray{1.94} & \lightgray{1.33} & \lightgray{1.15} & 0.01 & 0.02 & 0.01 & 0.02\\
    \hline

    \end{tabular}
    \caption{NMSE loss for downstream compression tasks using different pairs of training and testing datasets. When compression models are trained on real data and evaluated on generated data (Rows $1$,$2$) and vice versa (Rows $3$,$4$), our method records lower NMSE for corresponding real-generated dataset pairs, indicating that the data generated by our method is more similar to the real channel data. }
    \label{tab:cross_eval}
\end{table*}
\endgroup

\end{document}